\documentclass[11pt]{article}
\usepackage[version=4]{mhchem}
\usepackage[a4paper, margin=1in]{geometry}

\usepackage[T1]{fontenc}
\usepackage[utf8]{inputenc}
\usepackage{lmodern}

\usepackage{amsmath, amssymb, amsfonts}
\usepackage{siunitx}

\usepackage{graphicx}
\graphicspath{{Figures/}}
\usepackage{caption}
\usepackage{subcaption}

\usepackage{booktabs}

\usepackage{authblk}

\usepackage[numbers,sort&compress]{natbib}
\usepackage[colorlinks=true, linkcolor=blue, citecolor=blue, urlcolor=blue]{hyperref}

\title{A Heterogeneous 200 mm Silicon Nitride Photonics Platform for Visible-to-Near-Infrared Applications via Micro-Transfer Printing}

\author[1,2]{Konstantinos Akritidis}
\author[1]{Gaudhaman Jeevanandam}
\author[1]{Manuel Reza}
\author[1,2]{Maximilien Billet}
\author[1]{Jeonghwan Song}
\author[1]{Sandeep Seema Saseendran}
\author[1]{Vittal Prakasam}
\author[3]{Jan-Philipp Koester}
\author[3]{Jörg~Fricke}
\author[1,2]{Günther Roelkens}
\author[3]{Markus Weyers}
\author[1]{Roelof Jansen}
\author[1]{Pol Van Dorpe}
\author[1,2]{Bart~Kuyken}

\affil[1]{imec, Kapeldreef 75, 3001 Leuven, Belgium}
\affil[2]{Photonics Research Group, INTEC Department, Ghent University -- imec, 9052 Ghent, Belgium}
\affil[3]{Ferdinand-Braun-Institut (FBH), Gustav-Kirchhoff-Straße 4, 12489 Berlin, Germany}

\date{\normalsize Corresponding author: \texttt{Konstantinos.Akritidis@ugent.be}}

\begin{document}
\maketitle

\begin{abstract}
The commercialization of next-generation technologies, including optical interconnects, quantum computing, AR/VR, and medical diagnostics, requires a low-loss photonic platform offering compact, multifunctional systems in the visible and near-infrared range. Although silicon nitride (SiN) is an excellent material due to its ultra-low loss and broad transparency window, integrating active components such as light sources, modulators and photodetectors from diverse material platforms in a scalable, reliable way remains challenging. Micro-transfer printing is an emerging wafer-scale heterogeneous integration technology that can be implemented as a back-end post-processing step without disrupting the primary in-line fabrication process. In this work, we present a dual LPCVD SiN layer platform fabricated in a 200 mm CMOS pilot line, that incorporates micro-transfer printing modules, allowing the integration of active components on well defined recesses. A hydrogenated amorphous silicon layer is also available to increase the versatility of the platform allowing for evanescently-coupled III-V lasers as well as other passive functionality in the near-infrared region. We report full wafer-scale measurements showing low optical SiN losses of 4~dB/cm and 0.23~dB/cm at a wavelength of 488~nm and 940~nm respectively. In addition, a transition loss of only 0.35~dB is obtained from the SiN to the a-Si:H layer, in good agreement with simulated values. Finally, to showcase more advanced functionality, GaAs-based gain sections are micro-transfer printed on several dies, achieving consistent die-to-die lasing at 970 nm with on-chip optical powers of approximately 1~mW. These results showcase the potential of the integrated photonics platform towards unlocking a wide range of new applications in the sub-1-\textmu m spectral region.
\end{abstract}


\section{Introduction}

Leveraging mature complementary metal-oxide-semiconductor (CMOS) fabrication infrastructure together with advancements in hybrid and heterogeneous integration, silicon photonics (SiPh) has experienced remarkable growth \cite{liang_recent_2021,shekhar_roadmapping_2024}. By enabling compact, high-throughput, multi-functional systems-on-chip, SiPh has unlocked high-impact applications including optical telecommunications \cite{zhang_integrated_2026}, datacenter interconnects \cite{shi_silicon_2022}, LiDAR \cite{poulton_coherent_2017}, and sensing \cite{dhote_silicon_2022}. While commercial applications primarily operate in the O-band (1310 nm) and C-band (1550 nm), there is an ever-increasing demand to extend silicon photonics into the visible and near-infrared \mbox{(VIS-NIR)} spectrum. This submicrometre regime is vital for emerging fields such as AR/VR~\cite{shi_flat_panel_2025}, ion-trapping \cite{corsetti_integrated_2026}, atomic clocks \cite{newman_architecture_2019} and biomedical diagnostics \cite{song_visible_light_2025}. However, the 1.1 \textmu m bandgap of silicon results in prohibitive absorption losses in these ranges, which necessitates the adoption of alternative platforms.

\begin{figure}[!b]
\centering\includegraphics[width=1.0\textwidth]{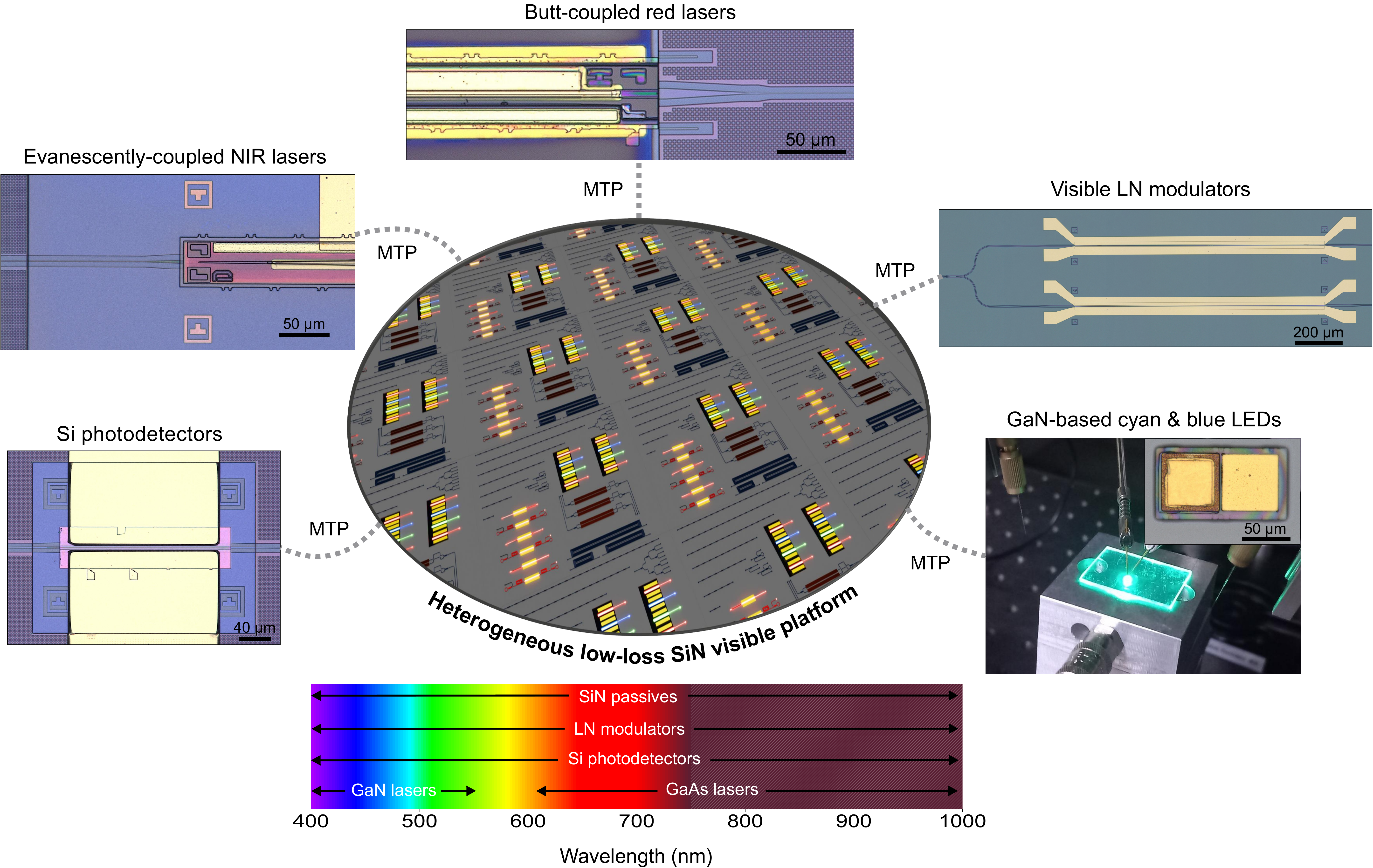}
\caption{Envisioned SiN platform offering a diverse active-passive library. The central schematic shows the conceptual integration layout, surrounded by optical microscope images of micro-transfer printed devices from ongoing developments. These include GaAs-based NIR and red lasers based on evanescent-coupling and butt-coupling, respectively, silicon photodetectors, thin-film LN high-speed modulators, and cyan- or blue-emitting GaN-based LEDs.}
\label{Roadmap}
\end{figure}

Among these materials, silicon nitride (SiN) has emerged as a premier candidate for photonic integrated circuits (PICs). Its widespread adoption is driven by its CMOS compatibility, wide transparency window, thermal stability, negligible two-photon absorption, and ultra-low propagation losses, capable of reaching sub-dB/cm levels at 405 nm \cite{morin_cmos_foundry_based_2021}. Furthermore, its moderate refractive index contrast relaxes strict lithographic constraints, offering significantly greater tolerance to fabrication imperfections and reducing scattering losses from sidewall scattering compared to silicon-on-insulator platforms. Crucially, low-temperature deposited SiN enables seamless Back-End-of-Line (BEOL) co-integration with CMOS electronics. This electronic-photonic convergence bypasses traditional packaging bottlenecks, paving the way for scalable architectures in co-packaged optics, high-speed interconnects and optical computing \cite{wan_integrating_2025}.

To date, various integration strategies have been explored to realize these platforms. For example, a foundry-fabricated visible SiN platform successfully incorporated flip-chip integrated 450 nm InGaN lasers alongside monolithically integrated silicon-based photodetectors \cite{mu_hybrid_2026}. This architecture utilized a butt-coupling scheme demonstrating 19 mW mean on-chip optical power with 3 dB/cm propagation losses at 445 nm. In the NIR regime, a complete SiN platform offering photonic building blocks, including amplifiers, lasers, modulators and photodetectors, was reported in Ref. \cite{tran_extending_2022}. In that work, wafer bonding was used to develop submicrometre-wavelength emitting tunable lasers reaching 10 mW output powers via a hybrid coupling approach. Specifically, the III-V semiconductor optical amplifier (SOA) was butt coupled to an intermediate dielectric layer and then evanescently coupled to the underlying SiN. This scheme differs from pure butt-coupling offering better alignment tolerances, while maintaining high coupling efficiencies. Furthermore, various other visible-light SiN platforms have been reported. These include architectures showcasing monolithically integrated silicon nitride-on-silicon waveguide photodetectors \cite{lin_monolithically_2022}, as well as platforms dedicated to the design and functionality of passive components \cite{romero_garcia_silicon_2013, sanna_sin_2024}, addressing key building blocks such as multimode interference (MMI) couplers, Mach-Zehnder modulators (MZI), microring resonators and waveguide crossings.

Collectively, these milestones underscore the significant interest in establishing a complete, high-performance visible photonics platform. However, despite the variety of heterogeneous integration technologies explored to date, a visible platform specifically tailored around micro-transfer printing (MTP) remains unrealized. MTP is a highly versatile integration scheme that seamlessly combines the high-throughput, parallel processing capabilities of wafer bonding with the known-good-die pre-testing advantages of flip-chip integration \cite{chen_micro_transfer_2026, roelkens_present_2024}. This technology has successfully enabled the integration of a wide variety of components, including photonic crystal cavities, lasers, modulators, photodetectors and light-emitting diodes across diverse material platforms such as lithium niobate (LN), indium phosphide (InP), lithium tantalate (LT), gallium arsenide (GaAs) and gallium nitride (GaN) \cite{murai_11_cm_long_2025, bommer_transfer_2025, haq_micro_transfer_printed_2020, niels_high_speed_2026, kiewiet_microtransfer_2026, chlipala_electrochemical_2025}. Ultimately, this technique facilitates the development of complex, multi-functional photonic circuits by selectively integrating these disparate materials. Crucially, it fully decouples the fabrication of the active components from the underlying SiPh foundry processes, which is an essential requirement for materials like LN that pose contamination compatibility challenges in standard CMOS manufacturing.

However, transitioning to shorter wavelengths presents a significant challenge for \mbox{SiN-based} photonics. The relatively low refractive index of SiN compared to III-V gain materials hinders the evanescent coupling scheme typically used in high-index silicon platforms, forcing a reliance on butt coupling. While butt coupling provides high coupling efficiency and effective heat dissipation, it suffers from significant misalignment sensitivity at shorter wavelengths. Furthermore, its reliance on facets, which are highly susceptible to degradation \cite{hempel_catastrophic_2010, wang_new_2021, tomm_mechanisms_2011}, requires specialized coatings \cite{nash_gaas_1979, ressel_novel_2005} or non-scalable cleaving processes that impede high-yield manufacturing. 

To achieve high yield, long operational lifetimes, and eliminate the need for optical facets, extending the reach of evanescent coupling to shorter wavelengths remains essential, necessitating the development of alternative strategies to overcome the refractive index limitations. One method involves fabricating III-V devices with ultra-narrow tips to achieve phase matching, enabling direct integration onto passive waveguides. For instance, Fabry-Pérot and distributed-feedback lasers emitting at 980 nm with powers exceeding 2~mW have been reported by integrating InGaAs-based SOAs featuring 100~nm wide tips directly onto tantala photonic integrated circuits via wafer bonding \cite{nader_heterogeneous_2025}. While this approach reaps many of the benefits of evanescent coupling, the reliance on such fine critical dimensions introduces significant fabrication complexity. An alternative solution that maintains CMOS compatibility with relaxed fabrication constraints involves introducing an intermediate layer with tailored optical properties. In our previous work, we demonstrated that employing a hydrogenated amorphous silicon (\mbox{a-Si:H}) interlayer with two etch depths enables phase matching, allowing for fully evanescently coupled lasers and amplifiers at 980 nm \cite{MY_REF}. However, while highly scalable in principle, this approach was previously demonstrated only at the sample scale using electron beam lithography and processes employed in an R\&D process environment. 

To bridge this gap and fully realize the MTP-compatible active-passive ecosystem introduced above, a transition to a standardized, 200 mm wafer-scale manufacturing flow is essential. By establishing a robust, CMOS-compatible SiN hosting platform, the integration of diverse active materials can be scaled from isolated lab demonstrations to a unified, multifunctional photonic library. Figure~\ref{Roadmap} outlines our long-term integration roadmap on this platform. Driven by rapid developments in micro-transfer printing  technologies for heterogeneous active integration, a diverse palette of material systems can be transferred onto the same SiN backbone. This includes, among others, GaN-based active elements for blue and green wavelengths, GaAs-based devices covering the red to 1000 nm range, thin-film lithium niobate for high-speed modulation, and silicon for photodetectors, addressing the broader challenge of spanning the entire visible-to-near-infrared spectrum.

As the foundational step towards realizing this roadmap, in this work we scale up and validate the core manufacturing and integration processes of this SiN platform. Specifically, we extend its functionality by integrating an a-Si:H layer module, leveraging the CMOS-compatible processing to achieve scalable, evanescently-coupled lasers emitting at the submicrometre wavelength range. We demonstrate consistent laser performance across several dies over two wafers, validating the robustness and uniformity of the evanescent coupling scheme. Additionally, we provide comprehensive wafer-scale characterization of critical passive components, including propagation and transition losses, and mirrors performance. The platform's versatility is further highlighted when considering the 800 nm butt-coupled continuous and mode-locked lasers previously demonstrated \cite{kiewiet_microtransfer_2026}. Together with the evanescently-coupled results presented here, the developed SiN platform offers a proven pathway for multi-wavelength, high-performance photonic integration.

\section{Design}

Developing a SiN platform that simultaneously covers the visible and near-infrared regimes presents a significant challenge. Spanning from 400 nm to 1000 nm, this range encompasses more than 1.3 octaves of bandwidth. To put this scale into perspective, an equivalent span in conventional telecom photonics would stretch from 1550 nm well into the mid-infrared at nearly 3900 nm. Managing such an expansive spectral window necessitates meticulous engineering of the passive waveguide stack in terms of both design and fabrication. 

\begin{figure}[!b]
\centering\includegraphics[width=1.0\textwidth]{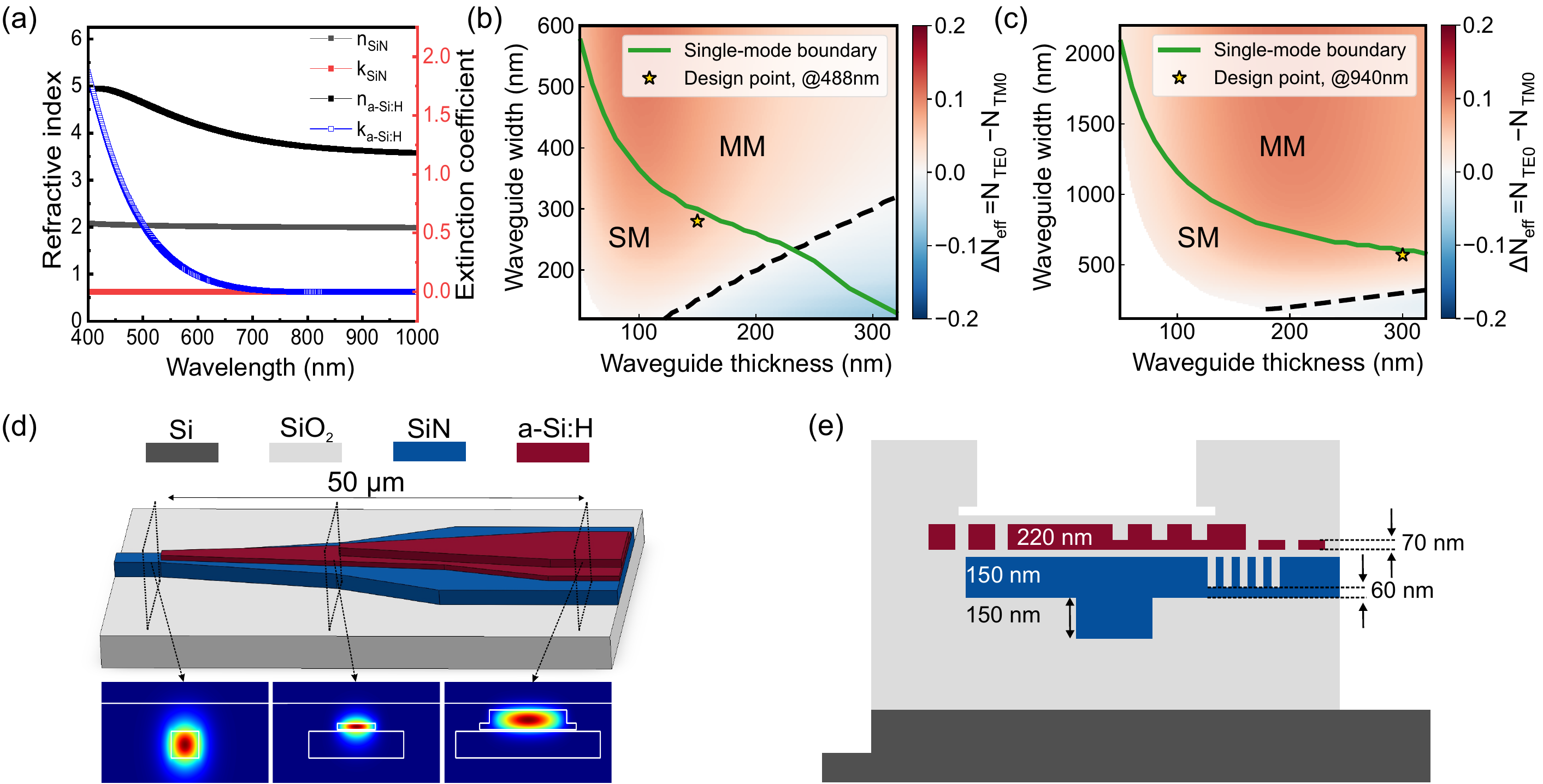}
\caption{Dual-layer \mbox{SiN} platform design and material characterization. (a) Ellipsometry measurements of the refractive index and extinction coefficient for the deposited \mbox{SiN} and \mbox{a-Si:H} recipes. (b-c) Simulated modal birefringence versus waveguide cross-sectional dimensions at wavelengths of 488~nm and 940~nm. The solid green curve defines the single-mode per polarization cutoff boundary (below which only the fundamental $\mathrm{TE}_0$ and/or $\mathrm{TM}_0$ modes are supported), the dashed line tracks the polarization-independent condition ($\mathrm{TE}_0$$=$$\mathrm{TM}_0$), and the star marks the selected operating point. (d) Schematic of the tapers designed for the adiabatic mode transition between the 300 nm thick \mbox{SiN} and the 220 nm thick \mbox{a-Si:H} waveguides. (e) Cross-section of the dual-layer SiN platform accommodating \mbox{a-Si:H} and micro-transfer printing modules for active device integration.}
\label{Concept}
\end{figure}

To identify material recipes yielding optimal optical properties, initially \mbox{SiN} and \mbox{a-Si:H} films were deposited on SOI wafers using various deposition parameters and subsequently characterized. Spectroscopic ellipsometry was first conducted to assess the refractive index and extinction coefficient. While this technique enables the assessment of macro-level material trends, it lacks the sensitivity required to accurately quantify the optical losses, necessitating complementary characterization techniques, such as prism coupling (Metricon) for moderate losses (roughly <~16~dB/cm) or the more versatile cutback method. Based on this screening, the optimal deposition recipes were selected for the integrated platform. The ellipsometry measurements of these final films are shown in Fig. \ref{Concept}(a). Notably, the fitted extinction coefficient of \mbox{SiN} remains zero across the entire wavelength range, whereas for \mbox{a-Si:H}, it drops to zero at approximately 790 nm. In reality, however, a baseline propagation loss of around 17 dB/cm at 970 nm is expected based on cutback measurements conducted via electron beam lithography from our previous work \cite{MY_REF}, which translates to an additional loss of only 0.17~dB for typical SiN-\mbox{a-Si:H}-III-V transitions with lengths on the order of 100 \textmu m.

To evaluate suitable \mbox{SiN} waveguide geometries, 2D finite-difference eigenmode (FDE) simulations were performed as a function of waveguide width and thickness using Ansys Lumerical. Simulations were executed at 488~nm and 940~nm to bound the extreme of the operating wavelength range. The resulting design spaces, shown in Fig. \ref{Concept}(b-c), map the geometric birefringence of the platform. The solid green curve marks the cutoff boundary below which higher-order modes are suppressed, maintaining a fundamental only ($\mathrm{TE}_0$ and/or $\mathrm{TM}_0$) regime. The dashed line tracks the polarization degeneracy condition ($\mathrm{TE}_0$$=$$\mathrm{TM}_0$). Together, these maps delineate the exact dimensional tolerances required to achieve single-mode per polarization operation while strictly maintaining $\mathrm{TE}_0$ as the fundamental guided mode.  Although thin and wide waveguides offer benefits such as lower propagation losses, reduced polarization sensitivity and relaxed lithography tolerances, a thicker waveguide geometry was selected to ensure higher optical confinement, more compact bend radii, and greater design flexibility. Consequently, out of the viable design space, a geometry with a 150 nm thickness and 280 nm width was selected for the blue wavelength regime, and a thickness of 300 nm and a width of 570 nm for the NIR. These design points are highlighted by the star symbols.

The choice of \mbox{a-Si:H} thickness is governed by the phase-matching condition required to enable efficient, adiabatic mode transitions at both the \mbox{SiN-to-a-Si:H} and \mbox{a-Si:H-to-III-V} interfaces. This requires matching the effective indices ($\mathrm{N}_{\mathrm{eff}}$) of the fundamental transverse electric ($\mathrm{TE}_0$) modes between the respective structures. While a thicker \mbox{a-Si:H} layer is essential to phase-match with high index (n $\approx$ 3.5) III-V SOAs, such a large thickness prevents efficient evanescent coupling to the lower index \mbox{SiN} layer. To resolve this trade-off, a dual-thickness \mbox{a-Si:H} scheme is implemented via the multi-stage taper configuration illustrated in Fig.~\ref{Concept}(d), which is accompanied by FDE cross-sectional profiles at critical locations along the transition. A 20 nm thick oxide layer resides between the layers to minimize perturbations to the mode. The entire transition spans a compact total length of only 50 \textmu m and consists of three distinct adiabatic sections. In the first section (25 \textmu m long), light from the 300 nm thick \mbox{SiN} waveguide transfers to a 70 nm thick \mbox{a-Si:H} layer; here, the \mbox{SiN} width linearly expands from 570 nm to 2 \textmu m while the thin \mbox{a-Si:H} layer tapers up from 150 nm, the critical dimension limit of our lithography, to 800 nm. In the second section (15 \textmu m long), the mode transitions to a 220~nm thick \mbox{a-Si:H} layer as the underlying \mbox{SiN} wides to 3 \textmu m, the thin \mbox{a-Si:H} expands to 900 nm, and the thick \mbox{a-Si:H} tapers from 150 nm to 500 nm. Finally, over the last 10 \textmu m, the thick \mbox{a-Si:H} waveguide expands to a width of 1.6 \textmu m to establish a high enough effective index to maintain confinement during the subsequent III-V integration, a value that can be readily adjusted depending on the specific active device requirement. Excluding material absorption losses, eigenmode expansion (EME) simulations performed in Ansys Lumerical indicate a total transmission close to 100\% when transitioning to the 70 nm thin \mbox{a-Si:H} and 98\% when transitioning to the 220 nm thick layer, thereby validating the adiabatic feasibility of the entire scheme without the need of very fine structures or complicated taper geometries.

Beyond facilitating the integration of evanescently-coupled lasers and amplifiers in the submicrometre wavelength range as previously reported \cite{MY_REF}, the availability of these two distinct \mbox{a-Si:H} thicknesses significantly expands the platform's versatility. For instance, the same architecture is well suited for integrated photodetection, since the photoconductivity of waveguide-coupled \mbox{a-Si:H} allows visible-light detectors to be seamlessly integrated on chip \cite{de_vita_amorphous_silicon}. Aside from active devices, these layers can be leveraged to enhance passive component performance; utilizing the higher absorption loss of \mbox{a-Si:H} at shorter wavelengths, thin gratings can be patterned on top of the SiN waveguides to selectively induce TM-mode loss, thereby improving polarization extinction ratios and simplifying component characterization. Finally, implementing porous \mbox{a-Si:H} structures provides an additional route towards on-chip chemical and biological sensing, further extending the platform's functionality.

The fully developed platform architecture is schematically illustrated in Fig.~\ref{Concept}(e). The buried oxide thickness is set to 2.9~\textmu m to prevent substrate leakage and minimize destructive interference from substrate reflections, providing an optimal trade-off for the different grating couplers across both the cyan and NIR regimes. In addition to the 150 nm and 300 nm \mbox{SiN} layer thicknesses, a 90 nm partial etch is introduced to yield a 60 nm thick \mbox{SiN} section. This thin layer can be utilized for spot-size expansion, to optimize fiber-to-chip or direct laser-to-waveguide coupling at the shorter wavelengths, as well as for evanescent coupling to thick, lower-refractive-index dielectric layers to implement hybrid butt/evanescent coupling schemes similar to the approach reported in \cite{tran_extending_2022}. Furthermore, localized recesses are defined above the \mbox{a-Si:H} and Si layers to accommodate evanescently-coupled and butt-coupled-based active components, respectively. Depending on the target application, these openings can be strategically positioned over the primary \mbox{SiN} waveguides to integrate modulators or photodetectors, establishing a flexible route towards full photonic functionality. 

\section{Fabrication}

\begin{figure}[!t]
\centering\includegraphics[width=1.0\textwidth]{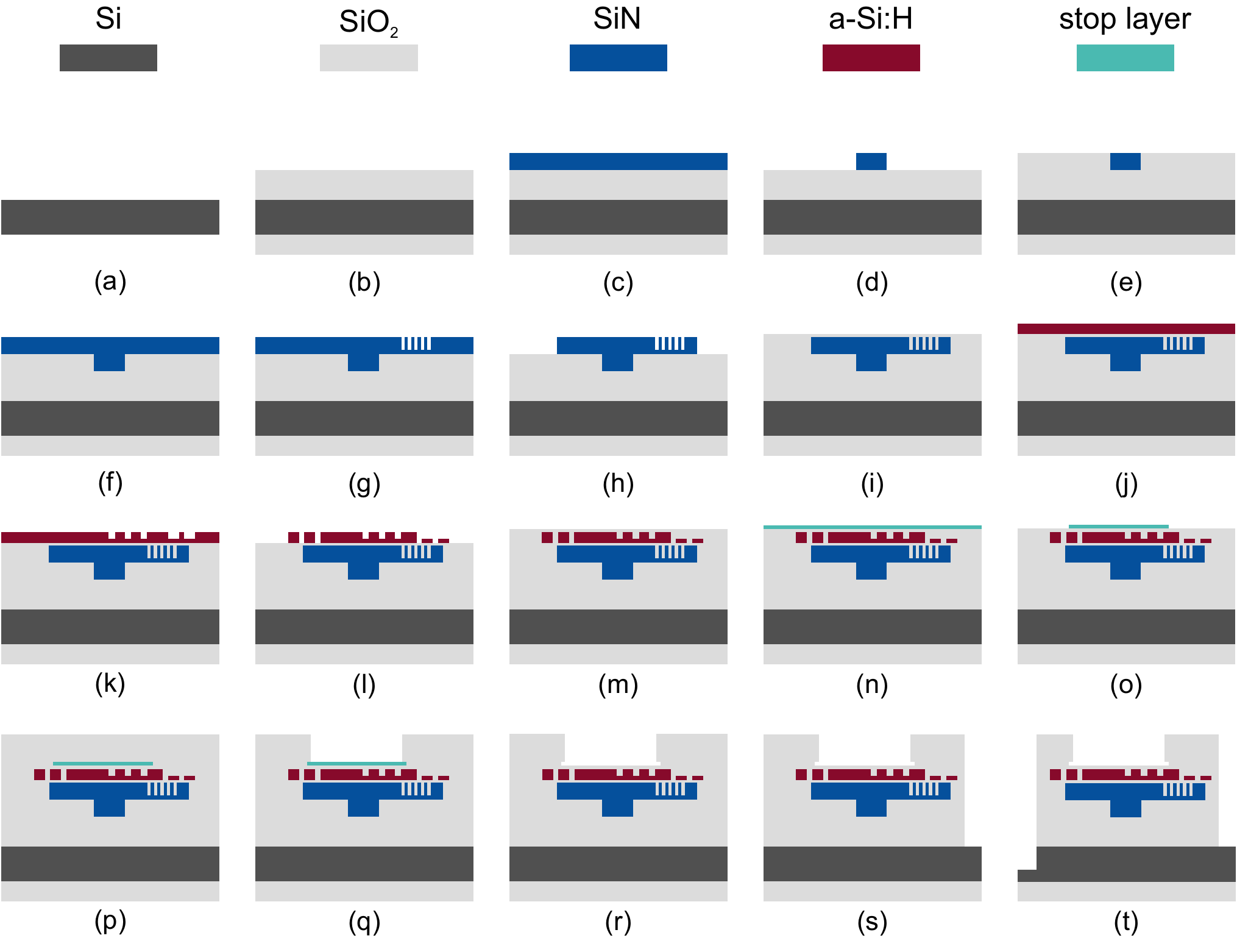}
\caption{Process flow for the fabrication of the platform. (a)~200 mm Si wafers. (b)~Deposition of 2.9 \textmu m bottom oxide cladding via thermal oxidation and HDP-CVD. (c-e)~Deposition of the first 150 nm LPCVD \mbox{SiN} layer, etching and planarization. (f)~Deposition of the second 150 nm LPCVD SiN layer. (g-i)~Partial etching of the top \mbox{SiN} layer to obtain 60 nm thickness, followed by full etching and planarization. (j-m)~Deposition of 220 nm \mbox{a-Si:H}, followed by partial etching to obtain 70 nm thick \mbox{a-Si:H}, full etching and planarization. (n-o)~Deposition and etching of the \mbox{aSi} stop layer. (p)~Deposition of 2 \textmu m top HDP-CVD oxide cladding. (q)~Etching to remove the oxide and landing on the stop layer. (r)~Selective etching of the stop layer. (s)~Etching down to silicon. (t)~Deep silicon etch.}
\label{Process_Flow}
\end{figure}

The wafers are fabricated on an industrial-scale 200 mm pilot line at imec with the process flow shown in Fig.~ \ref{Process_Flow}. Unlike custom, chip-scale laboratory prototyping, this wafer scale approach ensures high repeatability and yield required for commercial scalability. First, a 2.5 \textmu m thick thermal oxide is grown on the silicon wafer. This is followed by the deposition of 400 nm high-density plasma chemical vapor deposition (HDP-CVD) oxide, resulting in a total bottom oxide cladding thickness of 2.9 \textmu m. Subsequently, a 150 nm thick low-pressure chemical vapor deposition (LPCVD) \mbox{SiN} layer is deposited, patterned and etched. Afterwards, an oxide layer is deposited and planarized using the chemical mechanical polishing (CMP) technique to prepare for the second SiN layer. A second 150 nm \mbox{SiN} layer is then deposited, which is either fully and/or partially etched. By combining these two \mbox{SiN} layers, three waveguide thicknesses of 60, 150 and 300~nm are obtained, providing the flexibility required to successfully address the entire visible-to-near-infrared spectrum. With the planarization of the top SiN complete, a 220~nm thick \mbox{a-Si:H} layer is deposited via plasma-enhanced chemical vapor deposition (PECVD). This layer is then patterned using a two-step etching process to define the 70~nm and 220~nm thick regions. Following the etching process, oxide is deposited and planarized. Owing to the large topography and significant local mask density variations, maintaining a uniformly thin residual oxide layer across the wafer is highly challenging. Consequently, an 80~nm thick oxide layer was retained above the \mbox{a-Si:H} waveguides, to ensure sufficient protection. 


\begin{figure}[!t]
\centering\includegraphics[width=0.9\textwidth]{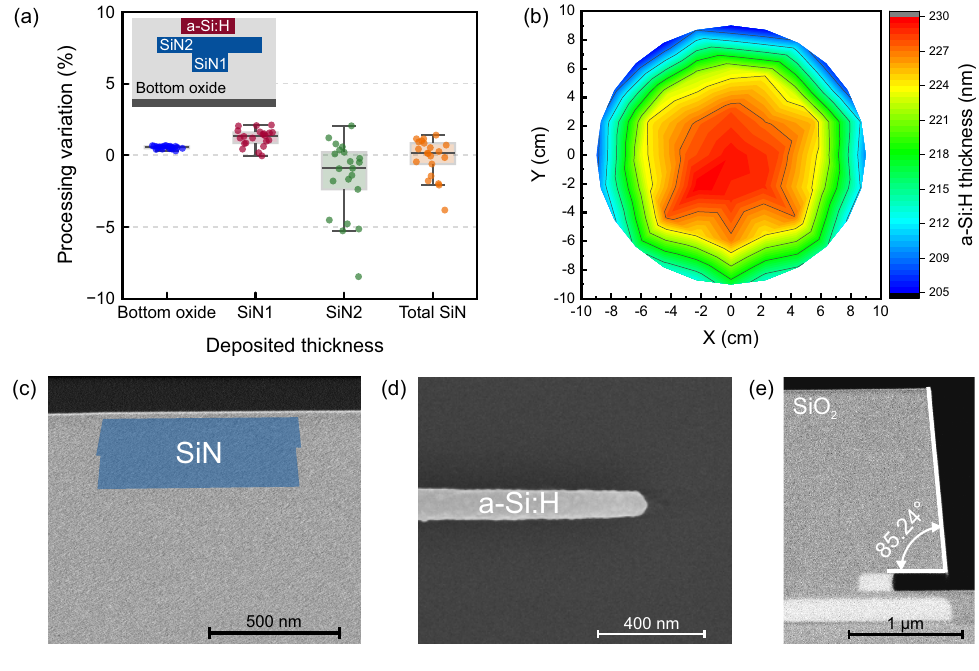}
\caption{(a) Processing variation of the deposited bottom oxide cladding and the dual-layer SiN across several wafers. (b) Colour-map showing the thickness distribution across the wafer for the nominal 220~nm thick a-Si:H layer. (c) Cross-sectional SEM of the dual-layer SiN (false-colored). (d) SEM of a representative test structure featuring the a-Si:H taper. (e)~SEM of the top oxide cladding recess with a measured sidewall angle of 85.24\textdegree.}
\label{Fabrication_Variations}
\end{figure}

To enable the integration of active components via micro-transfer printing, the platform includes trenches that expose the surface of the \mbox{a-Si:H} waveguides. To define these openings, an amorphous silicon (aSi) stop layer with good selectivity against oxide is first deposited and patterned, followed by a 2 \textmu m thick oxide top cladding deposition. To prevent unintended oxide etching caused by lithographic misalignment, the stop layer is designed slightly wider than the targeted opening regions (by 1 \textmu m on each side). After etching the top cladding locally, the stop layer is removed selectively exposing the \mbox{a-Si:H} still covered by the thin residual oxide. In applications requiring a reduced vertical separation between the waveguides and the active components, an additional etch step can be introduced. A second trench is also defined, extending down to the silicon substrate. This enables the integration of butt-coupled lasers while leveraging the superior thermal conductivity of silicon, which improves heat dissipation, allowing higher-power operation and reduced thermal loading. Finally, deeper trenches are also opened for fiber access and packaging.

To improve the wafer uniformity and reliability, dummy structures are utilized and key process parameters are monitored using a combination of metrology, scanning electron microscopy (SEM), and optical microscopy. The most important results are shown in Fig. \ref{Fabrication_Variations}. More specifically, in Fig. \ref{Fabrication_Variations}(a) the variations of the bottom oxide cladding and the deposition of the two SiN layers are shown. The bottom oxide cladding is very close to the target of 2.9 \textmu m, while for the deposition of the two SiN layers, excellent uniformity is also achieved, with most wafers falling well within the $\pm$5\% specification. In Fig. \ref{Fabrication_Variations}(b) a colourmap of the deposited \mbox{a-Si:H} is presented. The thickness obtained is larger in the center and decreases radially towards the edges. With a target deposition of 220 nm, the obtained range lies between 205 and 230 nm which correspond to -7\% and 2\% variation, respectively. Following critical processing steps, SEM images are taken for inspection. Fig. \ref{Fabrication_Variations}(c) shows a cross-sectional SEM of the two 150 nm thick SiN layers. A slight overlay misalignment is visible, resulting in a distinct profile; while this offset is minimal, future iterations could implement a thin oxide etch-stop layer or a wider top SiN design to mitigate this. In the case of the former, a reoptimization of the waveguide geometries would be required to maintain single-mode operation and efficient inter-layer transitions. In Fig. \ref{Fabrication_Variations}(d), an SEM of a representative test structure patterned on the 220~nm thick a-Si:H layer is shown. Although a narrower tip than the nominal critical dimension of 150 nm was fabricated, due to negative critical dimension bias during lithography and etching, this is desirable for minimizing insertion losses and reflection during inter-layer coupling. Finally, Fig.~\ref{Fabrication_Variations}(e) provides an SEM of the top oxide cladding opening, revealing a sidewall angle of 85.24\textdegree.

\section{Results}

\subsection{Platform Characterization}

\begin{figure}[!b]
\centering\includegraphics[width=1.0\textwidth]{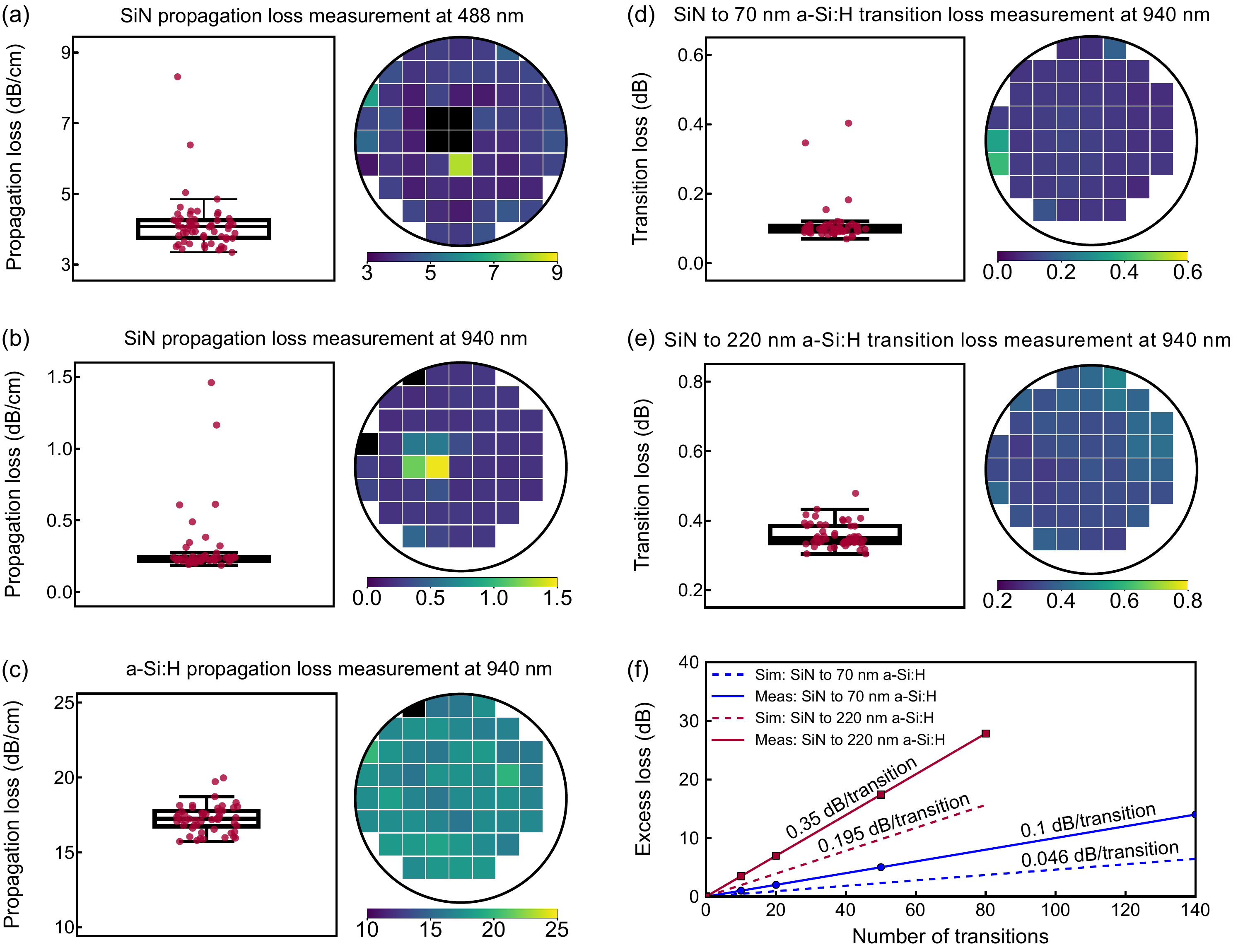}
\caption{(a) Wafer-scale propagation loss measurement of 150 nm thick SiN at a wavelength of 488~nm. (b-e) Wafer-scale performance characterization at a wavelength of 940 nm. (b) Propagation loss measurement of 300 nm thick SiN. (c) Propagation loss measurement of 220 nm thick a-Si:H. (d) Transition loss measurement for the coupling between 300 nm thick SiN and 70 nm thick a-Si:H. (e) Transition loss measurement for the coupling between 300 nm thick SiN and 220 nm thick a-Si:H. Box plots indicate the median and interquartile range (IQR), whiskers extend to 1.5$\times$IQR, and individual points represent die-level measurements across the wafer. (f) Comparison between simulated and measured excess loss for both SiN to a-Si:H transitions at a wavelength of 940 nm.}
\label{characterization_passives}
\end{figure}

For the assessment of the die-to-die variations and the devices' performance, wafer-scale measurements are conducted. The results are presented in Fig. \ref{characterization_passives} using a combination of box plots and colourmaps in order to provide statistical distributions and spatial insights. Black-coloured dies represent unsuccessful measurements, where a linear fit could not be obtained.

To extract the propagation losses of the different materials, the cutback method is employed by fabricating spiral waveguides of varying lengths connected with grating couplers used to couple light in and out of the chip. In the case of shorter wavelengths, 150 nm thick, 280 nm wide SiN spirals of lengths ranging from 1 to 9 cm are measured, using a commercial laser emitting at a wavelength of 488 nm (Fig. \ref{characterization_passives}(a)). From a total of 60 dies measured, a median propagation loss of 4.08 dB/cm is extracted; for 5 dies where the output signal at 9 cm fell below the noise floor, the shorter spiral lengths were used for the loss extraction. Notably, the maximum variation occurs at the center of the wafer, deviating from the typical edge-dominant variation patterns. This is attributed to the narrow bandwidth of the grating couplers. At shorter wavelengths, the reduced operating bandwidth exacerbates the impact of processing variations. Since a single-wavelength laser is used for the characterization, these spectral shifts cannot be calibrated out, resulting in higher observed variation at the center, where material thickness is greatest due to CMP. 

In contrast, for the characterization in the near-infrared band, a supercontinuum laser is used covering the spectral range from 900 to 1000 nm. The bandwidth of the grating couplers, patterned on the 300 nm thick SiN layer, is centered at 940 nm. Assuming negligible spectral variation over such short range, and hence a spectrally flat behavior, the losses are extracted by taking the median value over a wavelength window of 20 nm: $940 \pm 10$ nm. This method mitigates variations arising from their response, increasing the reliability of the measurements. For the characterization of the SiN, 300 nm thick, 570 nm wide spiral waveguides of lengths ranging from 1 to 9 cm are measured yielding a median propagation loss of 0.23 dB/cm (Fig.~\ref{characterization_passives}(b)). Out of the 51 dies, two dies located at the edges of the wafer could not be measured, while the largest variation was observed at the center following similar trend with the 488 nm measurements.

For the characterization of the \mbox{a-Si:H}, multimode spiral waveguides with a width of 1.6 \textmu m and lengths of 0.6, 1 and 1.4 cm, are fabricated on the 220 nm thick \mbox{a-Si:H} layer. These spirals are coupled to SiN waveguides through adiabatic taper transitions, designed for fundamental TE operation. While assessing the etching quality remains important, verifying the material's optical performance at shorter wavelengths is critical to the viability of the proposed coupling architecture. Therefore, a large width is chosen to minimize roughness-induced sidewall scattering losses, thereby isolating the contribution of material absorption. From a total of of 51 dies measured, presented in Fig. \ref{characterization_passives}(c), the propagation loss at a wavelength of 940 nm is found to be 17.22~dB/cm, with only one unsuccessful die located at the edge of the wafer. These results are in alignment with previously reported values \cite{MY_REF} consolidating the potential of this material for wafer-scale laser applications in the near-infrared band, where very short tapers are sufficient to couple the light from the SiN to the integrated III-V. 

The platform accommodates two distinct \mbox{a-Si:H} etch depths, catering to different application requirements. Consequently, the transition losses from the SiN to both \mbox{a-Si:H} thicknesses are characterized by measuring cascaded chains with an increasing number of transitions, with the results presented in Fig. \ref{characterization_passives}(d, e). Details regarding the design of the tapers can be found in the previous section. For the transition to the thin \mbox{a-Si:H} layer, across 50 measured dies, the median transition loss, which includes the propagation loss of all materials, is found to be only 0.11~dB/transition at a wavelength of 940 nm. In comparison, the transition to the 220 nm thick layer exhibits an increased median loss of 0.35 dB/transition, owing to the increased length required which contributes to both absorption and roughness-induced losses. Collectively, this platform achieves a 50\% reduction in loss compared to previous reports \cite{MY_REF} by leveraging mature foundry fabrication processes. 

Figure \ref{characterization_passives}(f) presents the experimental data alongside simulation results, which incorporate the \mbox{a-Si:H} material losses as extracted from the spiral waveguides measurements. The strong agreement between the two datasets validates the design, with the minor discrepancy attributed to roughness-induced scattering not accounted for in the simulations. These results demonstrate that despite the multi-stage nature of the transitions, uniform and high-efficiency performance is achievable across the entire 200 mm wafer, all while maintaining a minimum feature size of 150~nm, ensuring compatibility with standard lithographic processes.

\subsection{Integration of III-V amplifiers}

To extend the functionality of the platform and assess the reliability of the process, GaAs-based amplifiers are micro-transfer printed from the III-V source wafer onto SiN Fabry-Perot laser cavities across eight dies and two wafers. The dies' selection is targeting both central and peripheral positions to assess uniformity. The coupling scheme is based on evanescent-coupling through an \mbox{a-Si:H} interlayer which was previously reported on a sample scale using electron beam lithography \cite{MY_REF}. 
To ensure robustness against variations in the refractive index of \mbox{a-Si:H}, amplifiers with different taper transitions are fabricated \cite{koester_design_2025}. For the \mbox{SiN-to-a-Si:H-to-III-V} mode transition, an additional 100 \textmu m taper is appended to the previously described 50 \textmu m structure (Fig. \ref{Concept}(d)). In this taper the a-Si:H waveguide linearly decreases from 1.6 \textmu m width to 500 nm while the various III-V amplifier taper stages increase to adiabatically transition the mode. An additional 10 \textmu m taper follows narrowing the width of the interlayer down to 200 nm, essentially cutting off the waveguide and prohibiting energy transfer between the two structures. While the simulated transmission can reach beyond 90\%, amplifiers designed to transmit approximately 70\% (1.6 dB loss) from the 220~nm thick a-Si:H waveguide to the III-V SOA are chosen for the integration due to their increased robustness in negative variations of the refractive index of the a-Si:H layer. 

\begin{figure}[!t] 
\centering\includegraphics[width=1.0\textwidth]{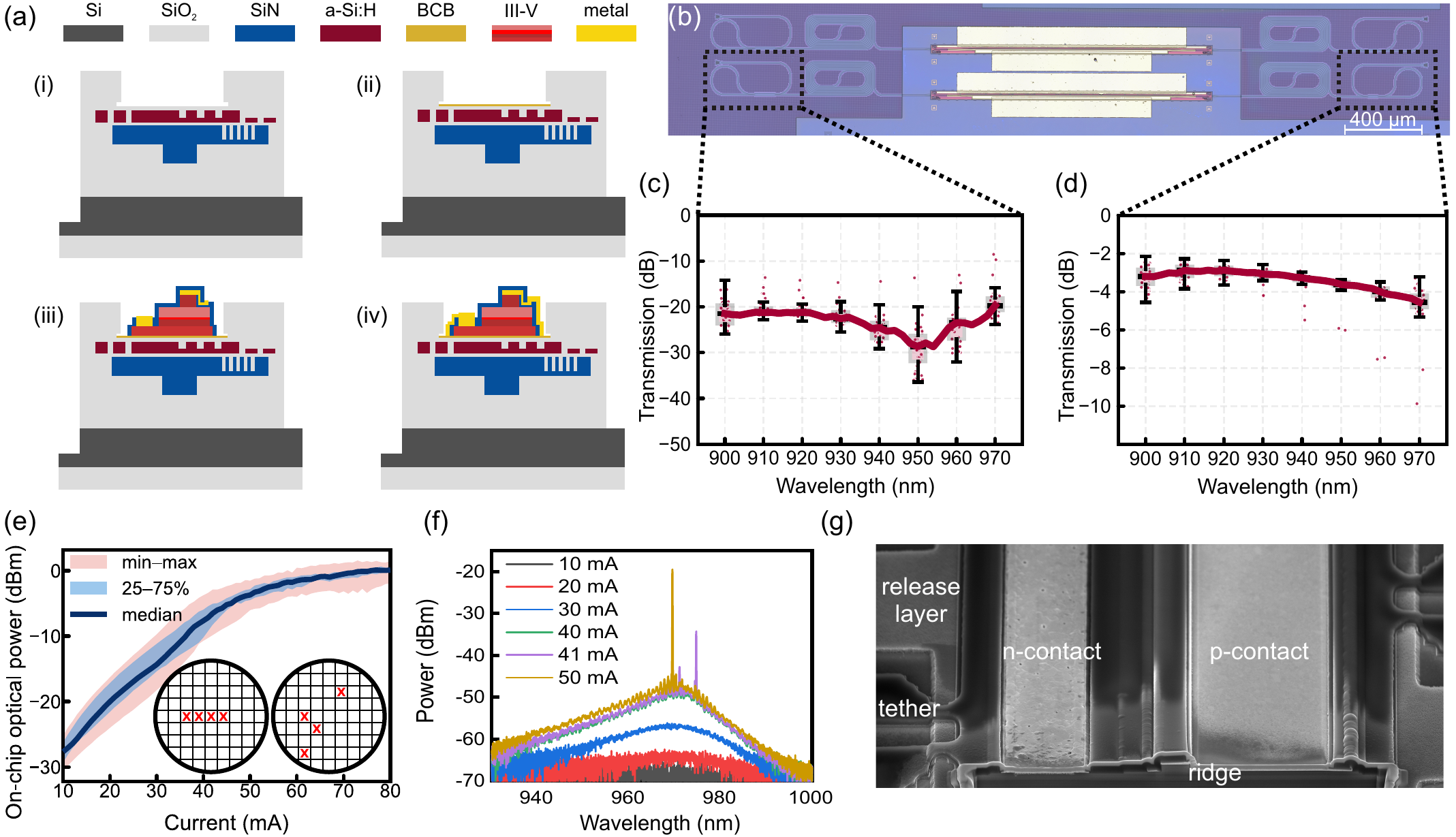}
\caption{(a) Process flow for the heterogeneous integration of amplifiers. (i) The passive waveguide platform following fabrication and dicing. (ii) Spin-coating of an adhesive BCB bonding layer. (iii) GaAs-based amplifiers micro-transfer printed onto the a-Si:H waveguides. (iv) Metallization for electrical contacting and characterization. (b)~Optical microscope image of the micro-transfer printed evanescently-coupled lasers. The SiN cavity consists of two Sagnac loop mirrors designed for 100\% and 50\% reflectivity, respectively. (c) Transmission spectrum of the high-reflectivity Sagnac loop mirror, with superimposed box plots at key wavelengths indicating the statistical distribution of performance across all dies. (d) Transmission spectrum of the output Sagnac loop mirror, featuring similar statistical overlays to demonstrate across-wafer uniformity. (e)~On-chip output power versus driving current for lasers from eight dies across two wafers. The median power at 80 mA is 1 mW. (f) Emission spectra at various injection currents, exhibiting lasing at a wavelength of 970 nm. (g)~Scanning electron microscope image of an amplifier on the source wafer prior to the underetching and transfer.}
\label{MTP}
\end{figure}

The process flow is detailed in Fig. \ref{MTP}(a). Following the initial fabrication, the wafers are diced and a short oxide blanket etch is performed on the dies to reduce the oxide layer on the exposed \mbox{a-Si:H} waveguides from 80 nm down to 20 nm. Afterwards, a thin benzocyclobutene (BCB) layer is spin-coated as an adhesive to facilitate the integration. The source wafer containing hundreds of amplifiers, is then selectively underetched using an \ce{NH4OH}:\ce{H2O2}:\ce{H2O} solution. Following the integration of the amplifiers, BCB curing takes place and contacts are metallized for measurements. While the specific integration of the amplifiers is performed at the die-level, this process is fully compatible with wafer-scale micro-transfer printers, which offer superior throughput and higher alignment accuracy. The wafer-scale capabilities of this heterogeneous integration technology has been already demonstrated for other active devices \cite{zheng2026microtransferprintinglithiumniobate}. An optical microscope image of the integrated GaAs-on-SiN lasers is shown in Fig.~\ref{MTP}(b). The amplifiers feature a length of 1.38 mm, with optical feedback provided by broadband Sagnac-loop mirrors designed for 100\% and 50\% reflectivity, respectively. Their transmission characteristics, evaluated from test structures across all dies (excluding one die at the wafer edge with abnormally high loss), are presented in Fig.~\ref{MTP}(c, d) for the 900 nm to 970 nm wavelength range. The high-reflectivity mirror exhibits -25 dB transmission at a wavelength of 940 nm, which corresponds to approximately 0.3\%. The observed die-to-die variation and spectral behavior are attributed to parasitic Fabry-Perot interference arising between the high-reflectivity mirror and the residual back reflections. Nevertheless, high reflectivities are maintained across nearly all dies, with transmission remaining below 5\% (-13~dB).  For the output mirror, a transmission of -3~dB is measured at 940 nm, showing excellent agreement with the simulations. Here, the largest die-to-die variation is observed at the wavelength extremities, due to approaching the cut-off bandwidth of the grating couplers. Collectively, these results, combined with the transition loss measurements reported previously, indicate highly consistent behavior of the SiN cavities across the entire wafer.

 To characterize the lasers, a single mode fiber coupled to the output grating coupler is routed via a splitter to an optical powermeter and an optical spectrum analyzer. This configuration enables the simultaneous acquisition of L-I (light-current) characteristics and emission spectra, respectively. To account for the grating coupler insertion loss, a reference SiN waveguide is measured on each die, providing the calibrated on-chip optical power. The results are illustrated in Fig \ref{MTP}(e, f). The lasers exhibit emission at around 970 nm with a threshold current of approximately $35 \pm 5$ mA and on-chip output powers of $0 \pm 1$ dBm at a driving current of 80~mA.  The devices exhibits a high above-threshold series resistance of approximately 28 \textohm, which is attributed to contact degradation during fabrication. Scanning electron microscope of the amplifier on the source wafer prior to the transfer process is shown in Fig. \ref{MTP}(g), revealing structural voids in the n-contact metallization. These holes allow the etchant to penetrate and attack the contact interface during the underetching step. This parasitic effect may be mitigated by employing a secondary metallization step to seal the voids or by depositing a sacrificial dielectric passivation layer to shield the contacts during the etch. These results represent an improvement over previously-demonstrated performance and, importantly, verify the high scalability and throughput associated with both the platform and the coupling architecture, making this approach ideal for mass production. These findings also suggest a path for optimization: by refining the amplifier design for higher transmission and optimizing fabrication to reduce the parasitic series resistance, higher powers are expected.

\section{Conclusion}
The development of a high-scalability SiN platform is a critical catalyst for advancing applications across the visible and NIR domain. In this work, we demonstrate a robust 200 mm wafer dual-layer SiN platform capable of seamless operation from the visible spectrum up to 1000~nm. By strategically incorporating a-Si:H and micro-transfer printing modules via specialized recesses, a versatile framework is established that supports diverse coupling architectures. A primary highlight of this versatility is the successful realization of evanescently-coupled GaAs-based lasers through an a-Si:H interlayer. This approach enables sub-micrometre lasing with output power exceeding 1 mW, showing good consistency across multiple dies. Coupled with highly uniform, wafer-scale low-loss propagation and transition performance, these results validate the platform's readiness for demanding high-yield applications.

Moving forward, we aim to establish a comprehensive photonic ecosystem by integrating modulators, photodetectors, and heterogeneous III-V or III-nitride lasers to achieve full spectral coverage. Future efforts will also focus on further reducing propagation loss, optimizing \mbox{III-V} device fabrication to increase output power and refining the integration of active materials. These advancements will pave the way for a fully integrated, high-performance photonic platform that meets the increasing demands of next-generation sensing, quantum information processing, and medical diagnostics.

\newpage

\section*{Acknowledgments}
We acknowledge funding from the Horizon Europe program of the European Union (Grant Agreement ID:
101070622). The support of colleagues in the Materials Technology, Process Technology, and Optoelectronics Departments at FBH for growth and processing of the III-V gain chips is gratefully acknowledged. Finally, we greatly appreciate Tangla David Kongnyuy and Aritrio Bandyopadhyay for conducting the wafer-scale passive measurements, as well as Valeria Bonito Oliva, Max Kiewiet and Tara Brstilo for providing some of the optical microscope images shown in Fig. 1.

\section*{Data availability statement}
The data that support the findings of this study are available from the corresponding author upon reasonable request.

\section*{Disclosures}
The authors declare no conflicts of interest.

\bibliography{references}

\end{document}